# The Temporal Evolution of Modality-Independent Representations of Conceptual Categories

Julien Dirani [*1] & Liina Pylkkänen[1,2,3]

[1]Psychology Department, New York University, New York, NY, USA, 10003
[2]Linguistics Department, New York University, New York, NY, USA, 10003
[3]NYUAD Institute, New York University Abu Dhabi, Abu Dhabi, UAE, 129188

julien.dirani@nyu.edu
liina.pylkkanen@nyu.edu

**Abstract**

To what extent does language production activate amodal conceptual representations? In picture naming, we view specific exemplars of concepts and then name them with a category label, like "dog." In contrast, in overt reading, the written word expresses the category (dog), not an exemplar.

Here we used a decoding approach with magnetoencephalography to address whether picture naming and overt word reading involve shared representations of semantic categories. This addresses a fundamental question about the modality-generality of conceptual representations and their temporal evolution. Crucially, we do this using a language production task that does not require explicit categorization judgment and that controls for word form properties across semantic categories. We trained our models to classify the animal/tool distinction using MEG data of one modality at each time point and then tested the generalization of those models on the other modality. We obtained evidence for the automatic activation of modality-independent semantic category representations for both pictures and words starting at ~150ms and lasting until about 500ms. The time course of lexical activation was also assessed revealing that semantic category is represented before lexical access for pictures but after lexical access for words. Notably, this earlier activation of semantic category in pictures occurred simultaneously with visual representations.

We thus show evidence for the spontaneous activation of modality-independent semantic categories in picture naming and word reading, supporting theories in which amodal conceptual representations exist. Together, these results serve to anchor a more comprehensive spatio-temporal delineation of the semantic feature space during production planning.

# INTRODUCTION

Concepts refer to internal representations that our brains hold about the world. We can arrive at those representations using different paths, such as recognizing a picture of a dog, or reading the word "dog." Intuitively, different routes to the concept "dog" lead to representations that are at least somewhat similar. However, the extent to which those representations actually overlap remains an open question. In other words, it remains unclear whether the concepts that the brain represents have a modality-independent component, or whether they are dependent on properties of the input stimulus. In addition, if amodal representations of concepts exist, it is unknown what their temporal evolution is and how it might interact with that of modality-specific representations.

Some theories hold that concepts are exclusively encoded via the perceptual system [1], with evidence indicating that processing conceptual information activates sensory-motor areas in the brain [2-5]. However, others have criticized this view for being too strong and proposed that while concepts have a grounding in perception, they also have an abstract component to them [6]. Bridging amodal and perceptual representations, the Hub and Spokes theory [7] describes an amodal semantic hub connected to modality-specific representations across sensory-motor areas (the spokes). Empirical evidence for shared representations across multiple modalities would support theories in which amodal concept representations exist, but obtaining such evidence faces various methodological challenges.

The investigation of the conceptual system in the brain is often done using picture stimuli and by comparing superordinate categories (e.g., animals, tools, faces). Previous work has indicated that pictures first activate low-level perceptual features within the first 100ms, followed by representations of semantic category (e.g. animal) within the first 150ms [8-12]. However, dissociating those categorical representations from low-level perceptual features in the picture stimuli remains a challenge, due to the systematic differences in visual shape across semantic categories. To remedy that, some studies have set out to investigate semantic category representations while controlling for the low-level features of the stimuli and showed that modality-independent representations are activated at around 100-200ms when viewing pictures and that they were sustained until 575ms [13-16]. These findings often use semantic category judgments or categorization tasks, and while this approach has proven to be valuable, it runs the risk of confounding neural representations of conceptual categories with neural processes associated with the selection of a category label for the experimental task. Thus, any findings of modality-independent representations across modalities could then be driven by a shared category-judgment process rather than an automatic activation of semantic representations.

Contrary to pictures, words do not explicitly contain information about semantic category membership at the perceptual level, but the temporal evolution of category representations using words is less well understood. For instance, there is disagreement about the latency of activation of semantic category representations using word stimuli. While some work indicates that semantic information is active between 200-500ms [14, 17], other findings argue for a 50ms locus of activation for spoken words [18] and 100-150ms for written words [19, 20]. Further, decoding superordinate categories using MEG data with words has proven to be difficult [16, 21], possibly because words do not necessarily activate superordinate categorical information as automatically as pictures do [22].

Overall, a fundamental question remains open about modality-independent semantic category representations during language production: Does a natural language production task with no explicit categorization automatically activate modality- and task-independent representations of semantic categories? Answering this question is crucial to understanding the nature of conceptual representations at the neural level and to shedding light on whether amodal conceptual representations exist, and if so, what their temporal evolution is. To address these questions, we measured neural activity with magnetoencephalography (MEG) during picture naming and word reading and analyzed the data using a decoding approach with generalization across time and modalities [23]. Generalization across modalities allowed us to investigate whether task-independent and sensory-independent representations of semantic categories were activated, while generalization across time assessed the temporal evolution of those hypothesized representations.

As a first hypothesis, it is possible that shared representations of semantic categories are activated rapidly after picture/word onset and develop into modality-specific representations at later stages of processing (H1). In contrast, it is possible that representations of semantic categories are activated in a bottom-up fashion, beginning as modality-specific, and later evolving into higher-order shared representations (H2). Regarding the temporal evolution of shared representations, if they are delayed in pictures compared to words, this would support a model in which semantic information is accessible from words ultra-rapidly, while in picture naming, non-perceptual representations of semantic categories appear at later stages of processing (H3). In contrast, if shared representations are delayed in words compared to pictures, this would support a model in which non-perceptual semantic category representations are activated in early stages of naming but in later stages of reading (H4).

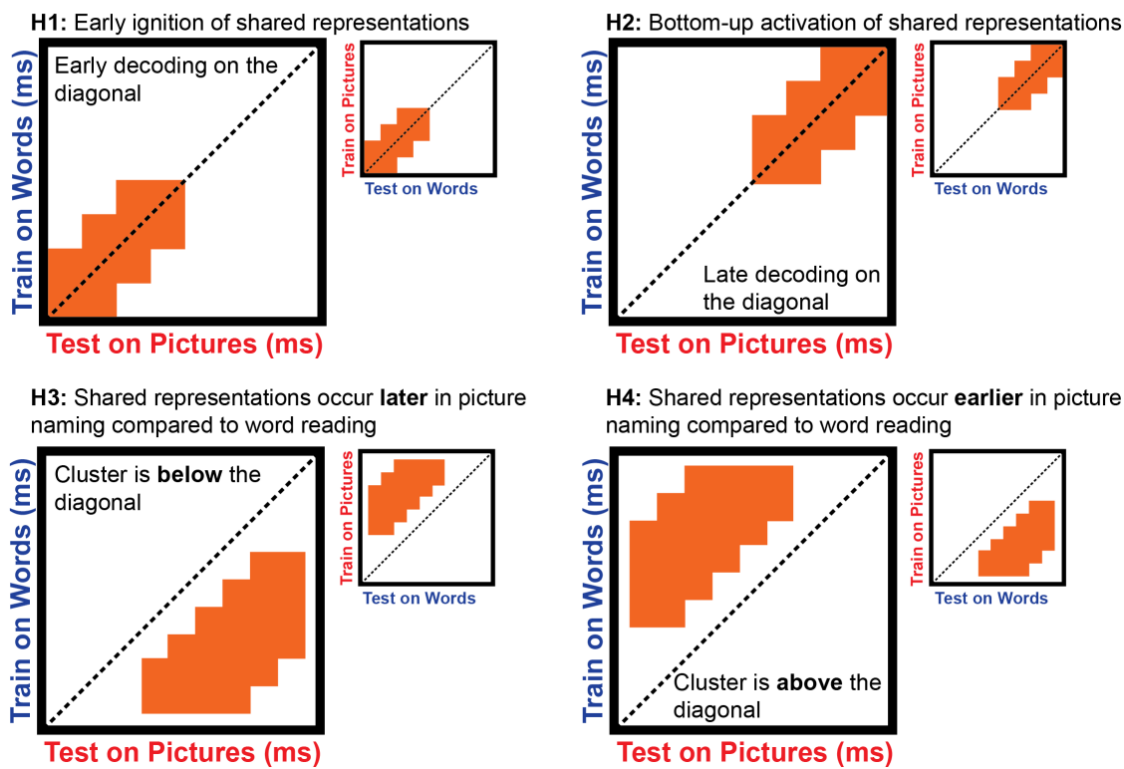

**Figure 1**: Hypotheses and their corresponding predictions. The large squares show the predicted patterns when training on words and testing on pictures while the small squares show the symmetrical pattern when training on pictures and testing on words.

## METHODS

### Participants

Twenty-four native English speakers were paid to take part in the study (14 female, Age: M = 24.75, SD = 4.63). All participants had normal or corrected-to-normal vision and reported no history of neurological or language disorders. The study received ethical approval from the institutional review board at New York University.

### Experimental design

The experiment required participants to overtly name pictures or read words out loud as they appeared on a screen while their brain activity was recorded using MEG. Prior to the experiment, participants went through a familiarization phase in order to make sure they would use the correct names for the pictures. Each trial started with a fixation cross that appeared on screen for 300ms, followed by a blank screen for 300ms, and finally the target picture or word appeared on screen for 1000ms (Figure 2). Participants were instructed to name the picture or read the word out loud as fast and as accurately as they could, and their responses were recorded using a microphone. The interstimulus interval length was sampled randomly from a normal distribution with mean and standard deviation both of 700ms. Stimuli were presented using Psychopy 2020.1.2 [24].

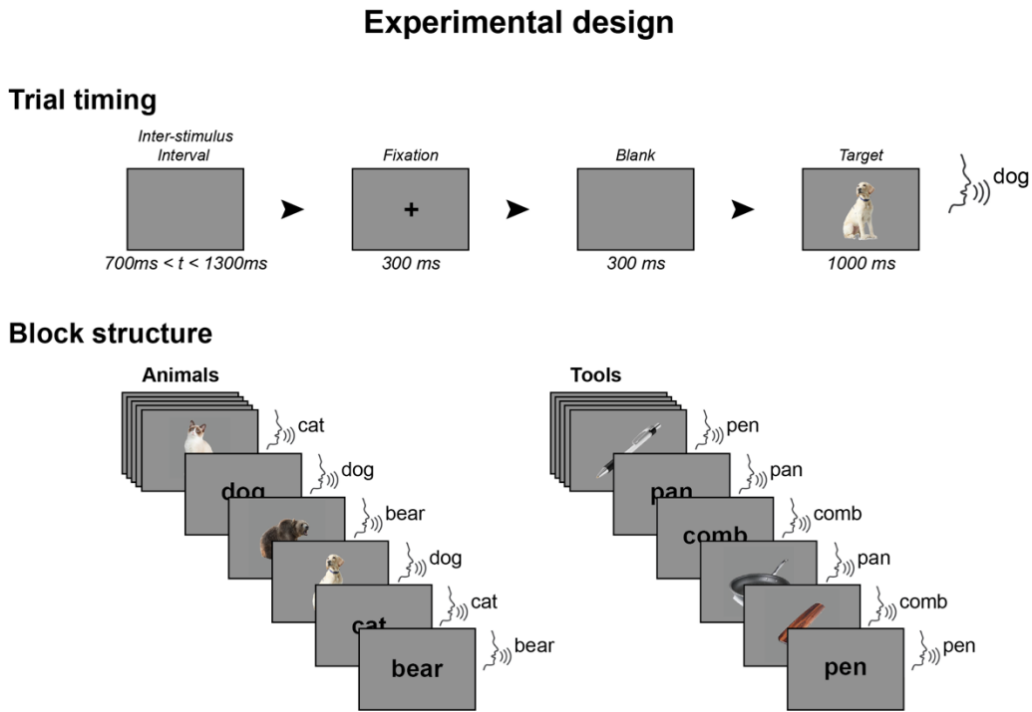

**Figure 2:** Experimental design

The stimulus set consisted of 50 unique exemplars from 2 semantic categories (25 animals, 25 tools). Each unique exemplar was repeated 10 times as a picture and 10 times as a word, for a total of 1000 trials. Trials were presented blocked by semantic category (animals or tools), while modalities (picture or word) were fully randomized within blocks. The names of exemplars were matched across the two semantic categories for number of letters, number of syllables, lexical frequency, concreteness rating, and number of morphemes (Table 1). Agreement as to the names of the pictures was obtained via the Amazon Mechanical Turk platform (www.mturk.com) where 50 participants were presented with the picture stimuli and were asked to report which single word they would use to describe each one. For each picture, the name that had the highest agreement between participants was selected. Finally, within each modality, both the size of the stimulus on screen as well as picture complexity were matched across categories. The size of the stimulus was determined by the ratio of foreground pixels to background pixels, and picture complexity was defined as the size of the picture in bytes.

|  | Animals | | Tools | | | |
| --- | --- | --- | --- | --- | --- | --- |
|  | M | SD | M | SD | t(24) | p |
| **Number of letters** | 5.08 | 1.75 | 5.16 | 1.52 | -0.17 | 0.86 |
| **Lexical Frequency** | 10472.40 | 13609.57 | 8162.76 | 18938.97 | 0.50 | 0.62 |
| **Number of orthographic neighbors** | 7.92 | 8.81 | 7.04 | 8.26 | 0.36 | 0.72 |
| **Number of phonological neighbors** | 14.80 | 15.44 | 13.92 | 13.02 | 0.22 | 0.83 |
| **Concreteness Rating** | 4.90 | 0.09 | 4.86 | 0.15 | 1.19 | 0.24 |
| **Average bigram count** | 3063.51 | 1527.16 | 3318.68 | 1328.78 | -0.63 | 0.53 |
| **Number of phonemes** | 4.16 | 1.40 | 4.08 | 1.15 | 0.22 | 0.83 |
| **Number of syllables** | 1.60 | 0.76 | 1.48 | 0.65 | 0.60 | 0.55 |
| **Number of morphemes** | 1.0 | 0.0 | 1.2 | 0.5 | -2.00 | 0.051 |

**Table 1:** Descriptive statistics and t-tests for lexical variables grouped by semantic categories (animals & tools). All $p > 0.05$ indicating that all variables were matched across categories. The statistics were taken from the English Lexicon Project [25].

## MEG acquisition and preprocessing

Continuous MEG was recorded with a 157-channel axial gradiometer system (Kanazawa Institute of Technology) at a sampling rate of 1,000 Hz with an online band-pass filter of 0.1–200 Hz. The raw data was noise-reduced with the continuously adjusted least-squares method [26] using the MEG Laboratory software 2.004A (Yokogawa Electric and Eagle Technology Corp., Japan). The data was low-pass filtered offline at 40 Hz and bad channels were identified after visual inspection, and the data for those channels were estimated using interpolation [27]. An independent component analysis was then fitted to the data using the "fastica" method, selecting components by 95 cumulative percentage of explained variance. Components related to eye-blinks, saccades, heartbeats, and flat channels were then rejected manually. Epochs from −100 to 600ms from target onset were extracted and baseline correction was done using the 100ms before the onset of the target. Time-locking of the epochs to the MEG triggers was adjusted using a photodiode. Epochs exceeding a maximum peak-to-peak threshold of ±3000 femto-tesla were removed automatically. Epochs corresponding to incorrect participant responses were excluded from the analysis. After all rejections, an average of 4.35 trials were rejected per participant. Evoked responses were created by averaging the 10 repeats of each exemplar. The resulting evoked responses were down-sampled by averaging bins of 5ms, and a principal component analysis (PCA) was performed to reduce the sensor space from 157 sensors to 70 components which explained at least 97% of variance for each individual subject. These steps were done in an attempt to increase the signal-to-noise [28] and resulting evoked responses were then used as the input for the decoding analyses described below.

**Analyses**

*Time-series decoding within modalities*

For each of the naming and reading modalities, a logistic regression classifier with l2 regularization was used to discriminate MEG response-patterns associated with each of the 2 semantic categories (animals and tools). The data was scaled so the mean activity at each feature (i.e. each PCA component) was 0 with a standard deviation of 1. A separate classifier was trained and tested at each time point from 0-600ms post stimulus onset for the words and pictures separately. Accuracy scores were obtained at each time point using a 5-fold cross validation. For each fold, the regularization strength of the classifier (C) was optimized using a grid search on a logarithmic scale between $1e^{-4}$ and $1e^{4}$ using a stratified 5-fold cross-validation. The regularization and the scoring were thus done on separate datasets. This procedure was done separately for each subject, and the average accuracy score across subjects are reported at each time point. In order to further investigate the organization of categorical representation over time, we also used the temporal generalization method [23] in which the training and testing procedure was repeated for all pairs of time points within each of the pictures and words modalities.

*Decoding with generalization across time and modalities*

In order to investigate whether and when modality-independent representations of semantic categories occur, we used a decoding approach with generalization across modalities and time. We used an identical approach to the one described above with the exception that the 5-fold cross-validation within modality was replaced by training the classifier on one modality and testing it on another. This procedure was done twice, once with the classifier trained on the words data and tested on the pictures data, and once where it was trained on the pictures data and tested on the words. This procedure was repeated for all pairs of time point and was done separately for each subject. In order to explore the spatial distribution of representations across the MEG sensors and over time, we plotted the model coefficients on the topographical sensor map for each of the training times. This was done by first reverse transforming the coefficient from the 70 PCA components back to the 157-dimensional sensor space.

*Group-level statistical testing*

In order to assess the timepoints at which decoding accuracies were above chance at the group level, we used a cluster-based permutation test. At each time point in the time series decoding, and each pair of time points in both the generalization analyses, a t-value was computed using a one-tailed one-sample t-test against chance (0.5). The resulting t-value map was thresholded at a t value corresponding to an uncorrected p-value of 0.05. Clusters were formed based on direct adjacency in time, and the sum of all t-values ($\sum t$) was computed for each resulting cluster. This procedure was then repeated by generating randomized data using random sign flips 10,000 times in order to obtain a null distribution. The Monte Carlo p-value was computed for each cluster in the original t map as the proportion of random permutations in which the observed ∑t was larger than the values from the permutation distribution. We retained clusters whose Monte Carlo p-value was smaller or equal to 0.05 [29]. The final results show the cluster of time-points where decoding accuracies were above chance at the group level. All analyses were performed using mne-python [30] and scikit-learn [31].

*Controlling for exemplar-specific and production related effects*

For our experimental task, participants were required to read the words and name the pictures out loud. The decoding method that we used aimed at linearly separating the superordinate categories in the MEG signal, thus the classifiers should, in theory, not rely on the single-trial information, which includes the production component of the task. In addition, since we have attempted to control for the linguistic properties of the words across semantic categories (Table 1), it is unlikely that the classifier relied on information at the word-form level.

Nevertheless, in order to empirically assess those assumptions, we ran the exact same analysis pipeline described above, with the exception that the MEG epochs were averaged over random subsets of exemplars within each of the "animals" and "tools" categories. In other words, rather than averaging over the same repeated exemplars, we average over bins of 10 random samples (Figure 3A, Analysis 2). The resulting evoked responses were then inputted into our analysis pipeline. This results in trials in which any information related to the individual exemplars is lost in the averaging process, allowing us to directly investigate whether the results found in the initial analysis can be explained by the production aspect of the task, or by information present in the word forms.

*Visual, lexical, and phonological variables*

In our main analyses described above, we investigated the temporal evolution and generalizability of conceptual representations by specifically looking at superordinate semantic categories. To offer a context for the semantic category findings, we conducted analyses assessing the activation of visual and phonological forms as well as lexical selection. We again used a decoding approach with generalization across time and across modality as described above. We operationalized visual complexity as the number of edges [32], phonological form activation with the number of phonological neighbors the word has and lexical access/selection as the log of lexical frequency. All word statistics were obtained from the English Lexicon Project [25]. There variables were split into binary bins of high and low values, with the splitting point being the median value of each word statistic, resulting in balanced classes. The remaining of the analysis pipeline was identical to what was described above for the classification of semantic categories.

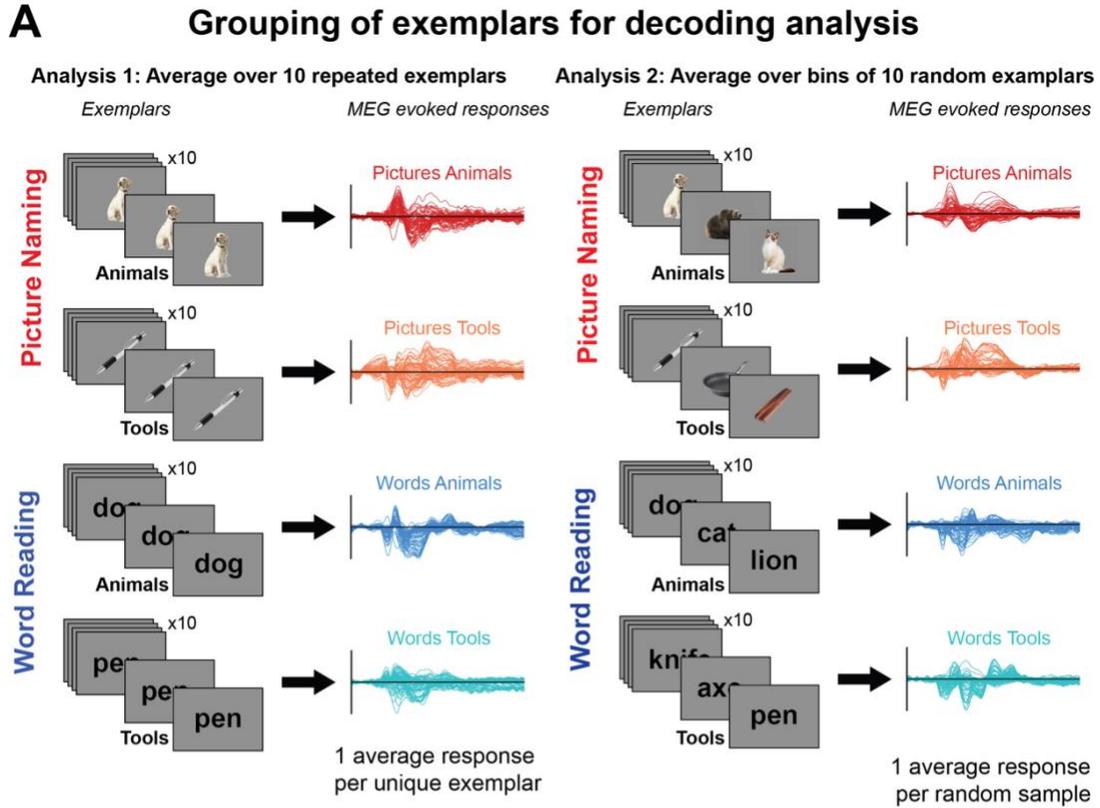

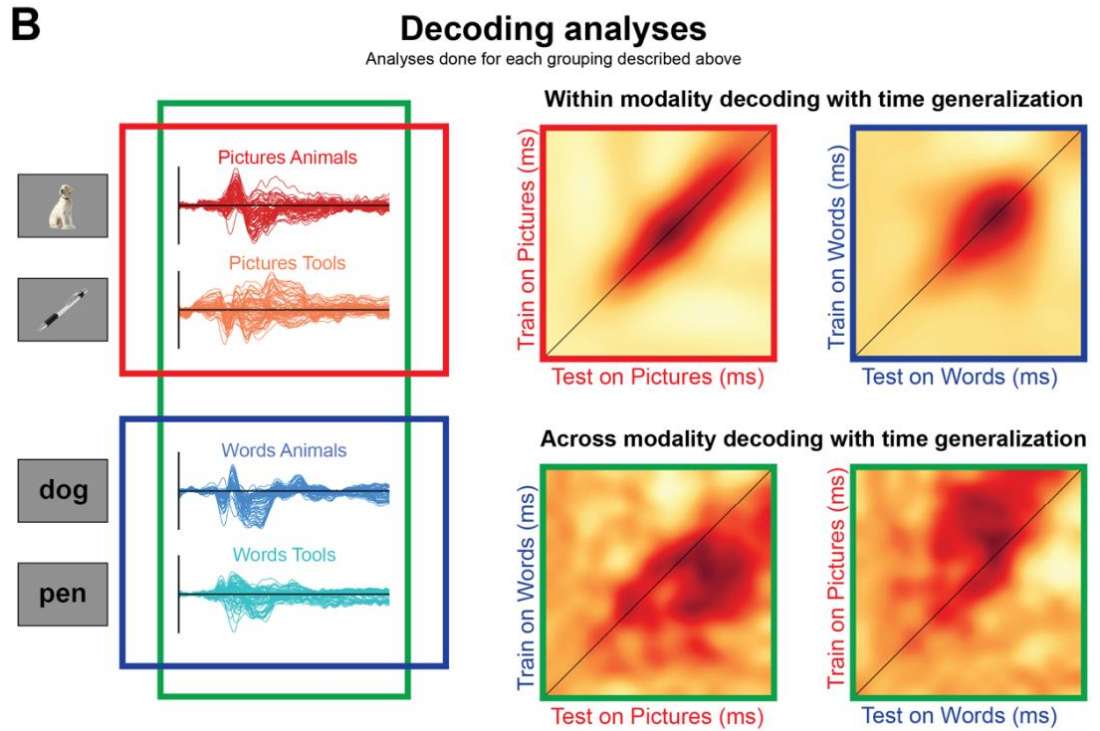

**Figure 3:** (A) Description of the two groupings of exemplars where first the averaging was done over all repeats of exemplars for each stimulus modality, and then averaging was done over sets of 10 random exemplars for each modality. (B) Analysis pipeline that was done once for each stimulus grouping described in (A).

# RESULTS

## Within modality decoding

The time-series decoding within modality yielded significant results for both word reading and picture naming. For the words, decoding accuracies were significantly above chance at 95-160ms and 180-455ms, with peak decoding at 350ms (62% accuracy). For the pictures, decoding accuracies were significantly above chance at 75ms-555ms, with peak decoding at 170ms (81% accuracy). While both modalities show an early and sustain activation of semantic category representations, peak accuracy was significantly higher in the pictures compared to the words ($t(46) = 5.13$, $p<.001$) . This not surprising given prior work also indicating that decoding categorical information elicited by words using electrophysiology has produced worse performances than with picture stimuli [21] likely because words do not necessarily automatically activate categorical automatically [22].

The word modality showed minimal generalization across time, suggesting that representations are consistently changing over time, indicating a feedforward process where representations are minimally sustained or reactivated at later times. The same pattern is observed in the picture modality, however, we also see a generalization across time for information activated at around 100-150ms, as indicated by the smaller cluster that is off the diagonal (Figure 4A). This could be due to visual information being sustained over time, or partially reactivated at later stages of processing, at around 300-600ms. The lack of symmetry across the diagonal can be explained by the fact that generalization scores are higher when the classifier is trained with high signal-to-noise data and tested with noisier data [23].

## Decoding with generalization across time and modalities

When training on the pictures and testing on the words peak decoding was reached at 415ms training time and 400ms testing time, with an accuracy of 0.57. On the accuracy matrices (Figure 5 A, B) we can see part of the cluster falls on and around the diagonal at 145-420ms. When training on the words and testing on the pictures peak decoding was reached at 380ms training time and 485ms testing time, with an accuracy of 0.57. The earliest time point where decoding across modalities was above chance occurred at around 150ms for both modalities. Notably, we observed no generalization across modality prior to 150ms, suggesting that representations of semantic categories prior to 150ms do not generalize across modalities are reflect modality-specific representations. This is complemented by the fact that the onset of within-modality decoding was at 75ms for pictures and 95ms for words (Figure 4), further supporting the interpretation that representations prior to 150ms are modality-dependent and most likely highly reliant on low-level visual information in pictures. Further, both matrices showed a large off-diagonal cluster indicating that representations around 150-400ms in words generalize at around 300-550ms for pictures. The projection of the model coefficients to the MEG sensor space indicated that early generalized representations localize to occipital areas and then later localize to bilateral temporal areas. Sensors in the occipital areas also indicated high values at around 400ms (Figure 6).

**Controlling for exemplar-specific and production related effects**

In order to demonstrate that the results previously described where not driven by exemplar-specific word form representations or by production-related motor planning, we repeated the exact same analysis pipeline with the exception that evoked responses were created by randomly sampling bins of 10 random exemplars for each modality, and then averaging the MEG data within each bin. The results revealed a strikingly similar pattern to that of the original analysis (Figures 4B, 4C, 5B, 5C) suggesting that our findings are unlikely to be driven by exemplar-specific variations or by word-form representations. A couple of differences were nevertheless observed. First, for words, the within-modality decoding with time generalization indicated an earlier and more sustained representation of semantic categories in the random-exemplar grouping compared to the repeated-exemplar grouping. Second, the across modality decoding indicated that when training on the words, the portion of the cluster falling on the diagonal is no longer significant (Figure 5C) however, when training on the pictures, the early diagonal cluster remained significant ($p<.05$; Figure 5D).

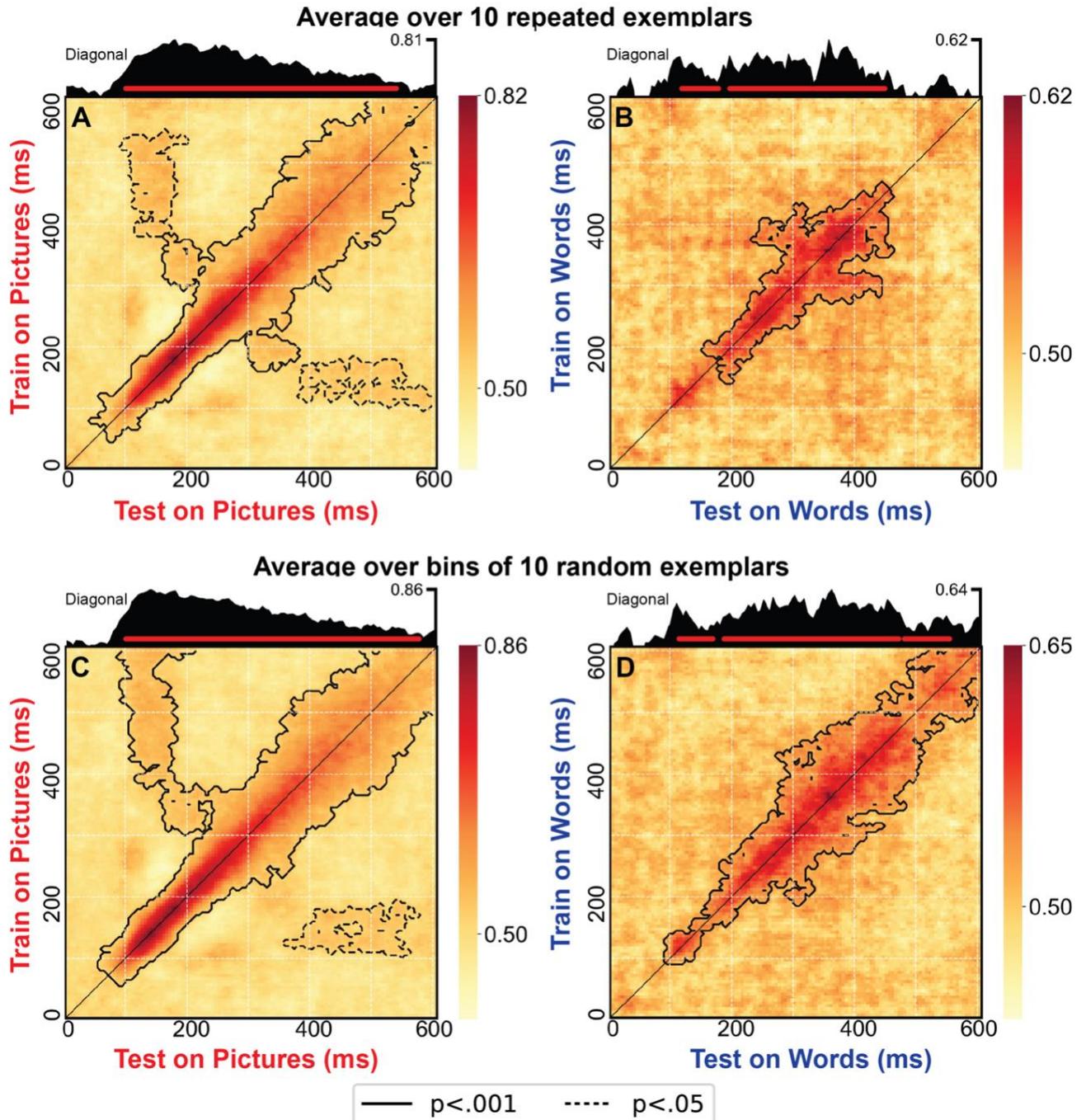

**Figure 4:** (A-B) The decoding results show little generalization for the words, and a generalization at ~150ms for pictures to ~300-600ms. The shaded plots above the matrices indicate the accuracy scores at the diagonal with decoding onsets of 75ms in images and 95ms in words. (C-D) Results of the second analysis with averaging over sets of 10 random exemplars for each modality. The patterns of results are similar to the first analysis suggesting that the decoding results are unlikely to be driven by exemplar-specific representations or by the word forms. Contour plots indicate clusters of time-point pairs with accuracy scores significantly above chance.

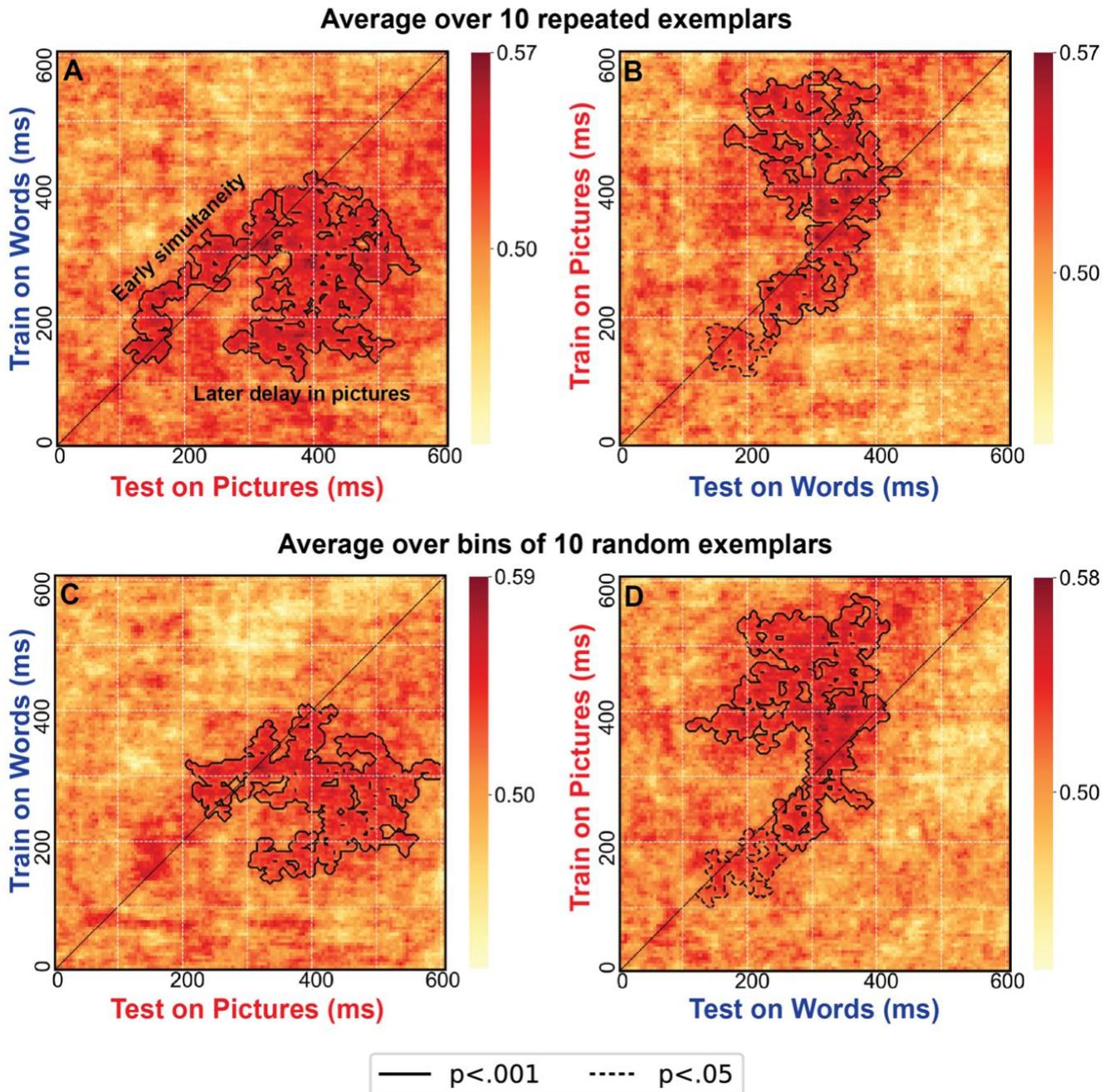

**Figure 5:** (A-B) Decoding across modality is significant around the diagonal earlier starting at ~150ms and then delayed in images compared to words starting 400ms supporting hypotheses 1 and 2. (C-D) Results of the second analysis with averaging over sets of 10 random exemplars for each modality. The patterns of results are similar to the first analysis suggesting that the decoding results are unlikely to be driven by exemplar-specific representations or by the word forms. Contour plots indicate clusters of time-point pairs with accuracy scores significantly above chance.

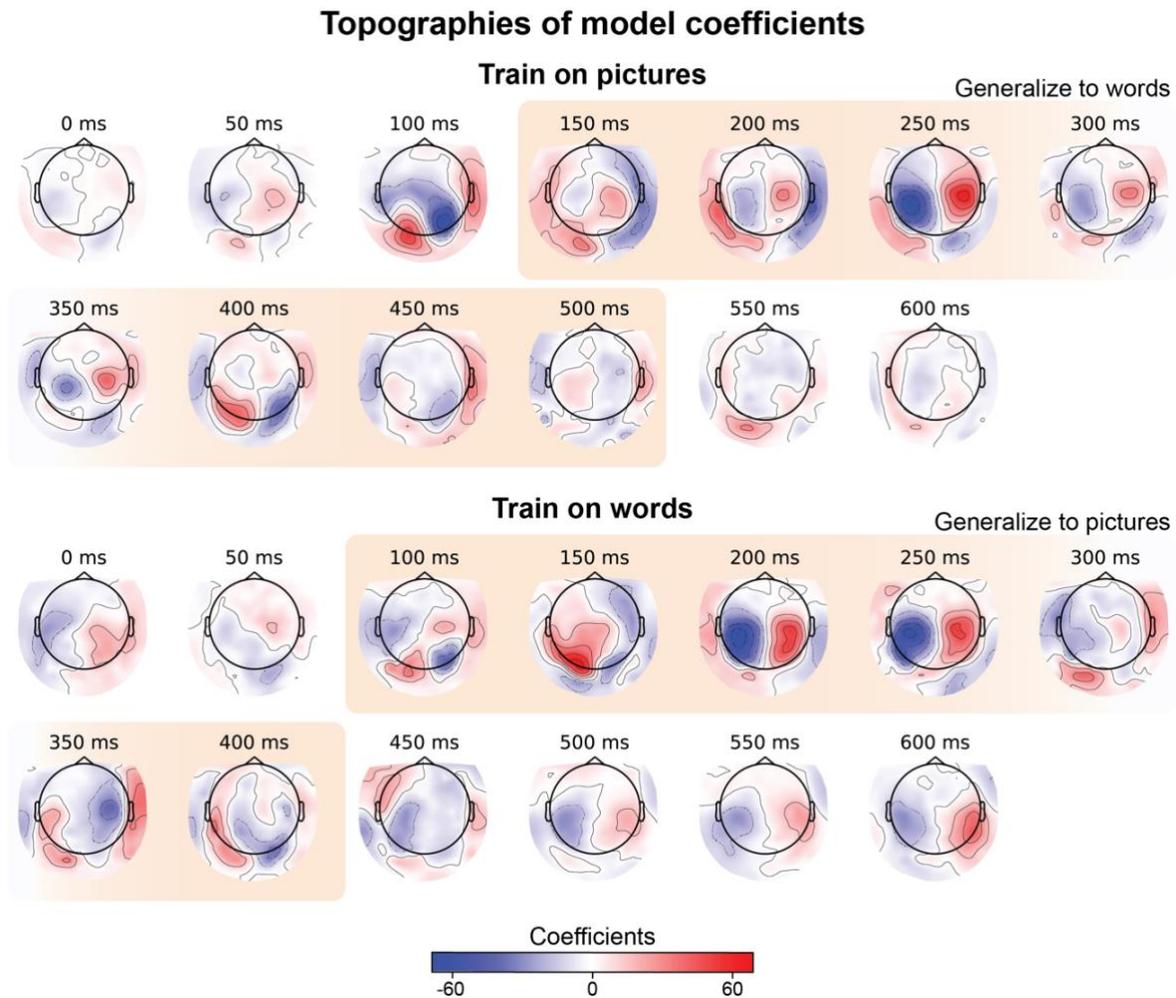

**Figure 6:** Projection of the model coefficients on the MEG sensor topographies showing that modality-independent representations localize to occipital areas first and later to bilateral temporal areas. The orange shaded areas indicate the time windows at which the classifiers generalized across modalities.

## Interaction of semantic categories with visual, lexical, and phonological variables
*Within modality decoding*

In picture naming, decoding onset of visual representations occurred very early at 60ms post stimulus onset, followed by phonological representations at 90ms, and finally lexical representations at 115ms. All of the visual, semantic, and lexical representations showed a sustained activation, with the exception of phonological representations which were classified with above chance accuracy in the 100-150ms and then in at 400-500ms. Notably, the later 400-500ms time window only reached significance in the analysis with generalization across time and not in the time-series decoding (Figure 7A). Further, all variables with the exception of semantic category showed little generalization across time, as indicated by the clusters falling on and around the diagonal. In word reading, decoding of visual, lexical, and phonological representations started respectively at 60ms, 80ms, and 70ms, indicating that those representations are activated virtually simultaneously, contrary to the serial pattern that was observed in picture naming (visual > phonological > lexical) (Figure 7B). Since phonology is being operationalized by number of phonological neighbors, the serial pattern observed in picture naming could represent the activation of phonologically similar words prior to convergence on a target word. This possibility is further considered in the discussion section. Finally, the generalization across time shows that visual and phonological showed a sustained activation until ~400ms, while lexical representations were active at 80-125ms, and then later at 250-400ms. Notably, those timings of lexical activation are in line with previous work suggesting an initial rapid access to lexical information as well as a later lexical access stage [17, 19, 20].

*Across modality decoding*

The results of the decoding across modality with time generalization (Figure 7C-D) revealed two notable patterns. First, for both the visual and phonological variables, the classification generalized across modalities early and simultaneously at ~80-125ms, reflecting an early locus of modality-independent representations of lower level visual and phonological features. Second, shared representations of lexical frequency occurred at 100-300ms in words, corresponding with the 400-500ms time window in pictures, in line with models of word production which predict faster lexical access in reading compared to naming [33].

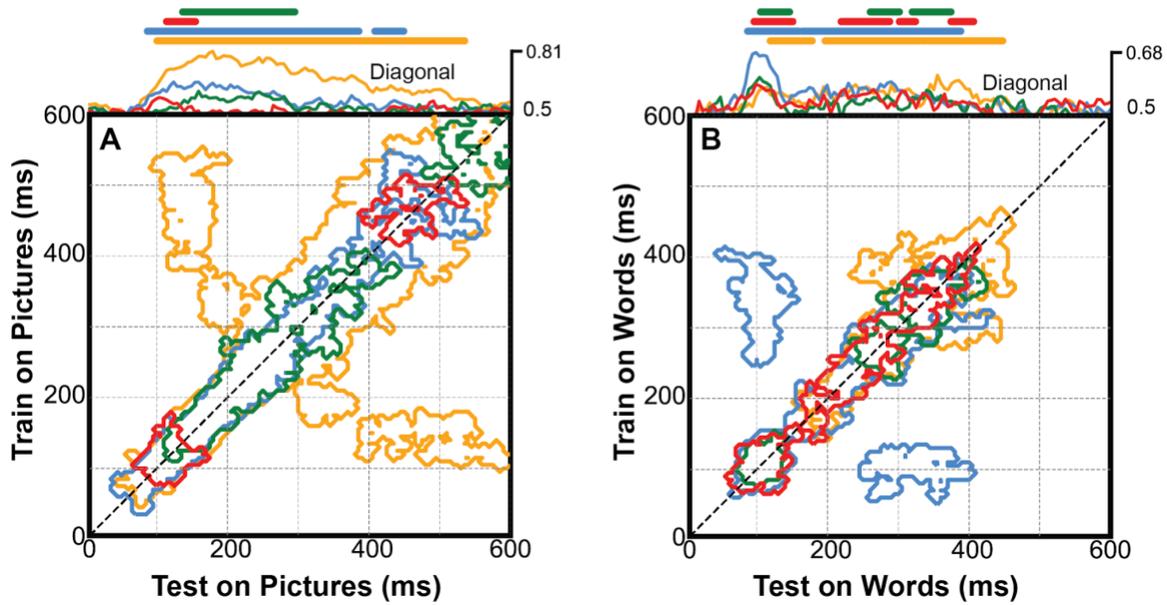

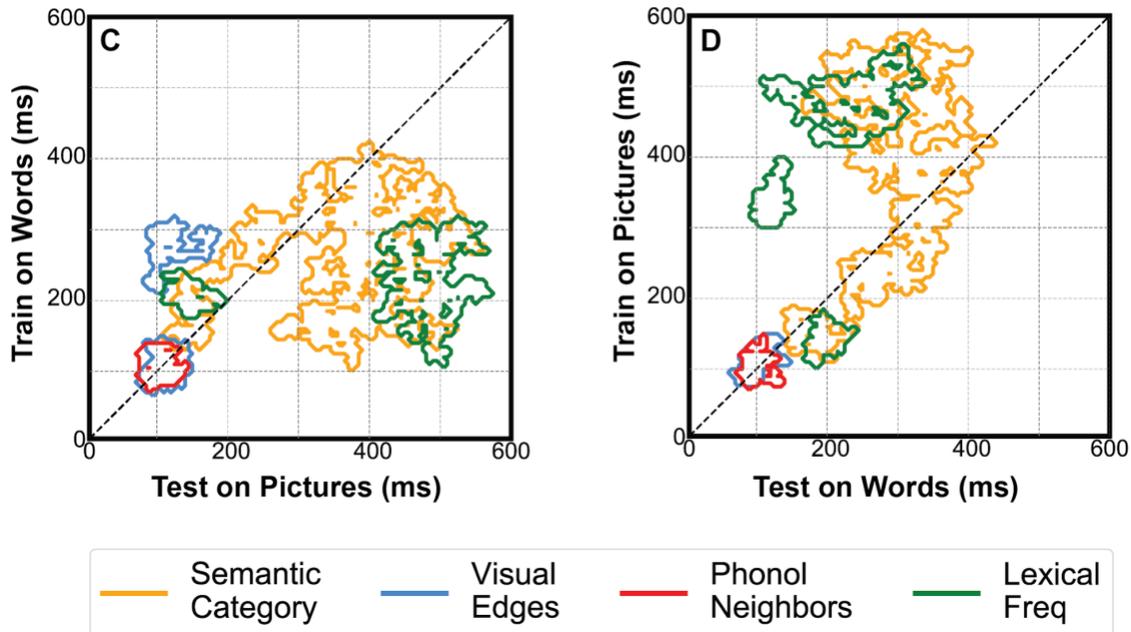

**Figure 7:** Contour plots indicate clusters of time-point pairs with accuracy scores significantly above chance for each of the 4 variables. The line plots above the matrices indicate the accuracy scores at the diagonal. (A) Category representations are activated virtually simultaneously with visual representations in naming, (B) but after lexical access onset in reading. (A-B) Picture naming shows seriality in decoding onsets of visual, phonological, and lexical representations compared to virtually simultaneous onsets in reading. (C-D) Visual and phonological representations generalize early and simultaneously, while lexical representations are delayed in naming in line with models of word production [33].

# DISCUSSION

While previous neuroimaging findings have provided some evidence for the existence of modality-independent representations of semantic categories [16, 34-38], it has remained unclear whether those representations are automatically activated in language production in the absence of explicit categorization judgments. We combined the high temporal resolution of MEG with a decoding approach with generalization across time and modalities [23] to assess the timing of those hypothesized representations in picture naming and overt word reading. We also compared the temporal evolution of categorical information within each of these two modalities. Our results indicate that both in picture naming and word reading, modality-independent representations of semantic categories are automatically activated slightly later than their respective modality-specific representations. The modality-independent representations first evolved in parallel across both modalities and were then delayed in picture naming compared to word reading. We demonstrated empirically that these results are unlikely to be driven by exemplar-specific representations, word forms, or motor processes related to the production of words. Finally, we also provided evidence for the spatial distribution of semantic category representations, as well as their interaction with visual, lexical features of the target concept. Below we first discuss the implications of our findings for each specific modality before addressing the modality-independent findings in more detail.

## Modality-specific representations

*Picture naming: Semantic categories active at 75ms*

Our results indicated a mostly feedforward evolution of semantic category representation in picture naming starting at 75ms. In line with previous findings [39], pictures showed a quick evolution of categorical information over time indicating that representations are rapidly updated. As information evolved over time, generalization across time appeared to become more pronounced, suggesting that representations are sustained for longer at later stages of processing [9], or that multiple information processing steps are gradually accumulated over the processing pipeline [39]. We also found that representations active at around 100-150ms emerge again at around 300-600ms post stimulus onset. One previous study had found a similar pattern using an experiment where participants passively saw pictures without a language production component [40]. This generalization effect could be due to visual information being sustained over time, or partially reactivated at later stages of processing (~300-600ms). In fact, the 100ms time locus has been associated with the activation of low-level perceptual features in picture-naming tasks [12] making it possible that the generalization of representations at this time to the later 300-600ms time window indicates a reactivation of perceptual features later on in the timecourse of picture naming. Nevertheless, evidence also showed that modality-independent representations of semantic category are activated as early as ~110ms [14, 16], which prompts the possible alternative explanation that the early 100ms window combines low-level features as well as modality-independent representations and is reactivated later or sustained over time.

*Word reading: Semantic categories active at 95ms with minimal generalization across time*

In the word modality, the 95ms decoding onset of semantic category supports the hypothesis that semantic representations are accessed rapidly when reading words and is line with prior work that used spoken words [20]. While other findings pointed to a later onset of semantic category access at around 200ms using words [16], this discrepancy can be due to the use of a different experimental task, since those results were found in the context of a categorization task rather than a language production task and using stimuli of faces and places as opposed to animals and tools as used here. In our results, peak decoding was only reached at around 350ms in the word reading modality confirming the intuition that representations of semantic categories are more strongly present after the surface level properties of the word have been processed and lexical access has occurred [17, 19, 20]. Interestingly, the time-generalization results revealed that the activation of semantic categories from word stimuli follows a feedforward process with virtually no generalization over time. Marginally more generalization occurred later in the timecourse, at around 400ms, where decoding accuracies cluster further away from the diagonal; however, the off-diagonal cluster remains too small to implicate thorough theoretical conclusions. While in the picture modality, this feedforward evolution can be explained by the fact that picture naming is a process that starts with the recognition of the picture based on low-level visual features which rapidly evolves over time and accumulates evidence to reach a higher-level representation of semantic category [39, 41], the parallel observation when using words is novel. Models of word recognition typically describe the process as a mapping from letters to phonemes and finally to motor commands [33, 42]. Semantic access is described as occurring in parallel, but little is known with regards to the specifics of the conceptual representations that become active during this time. Here our findings suggest that at the category level, representations spontaneously elicited by words seem to be continuously updated over time and peak at around 350ms. One possibility is that different stages of processing activate semantic category information at different level of granularity, or different types of knowledge associated with a category. To illustrate, tools could activate representations of motor commands, non-living things, inanimate objects, or more broadly, a common context in which tools are more likely to be observed. Those representations could systematically be activated at different stages of processing and evolving over time, which could be reflected in a decoding pattern such as the one that we observed here. More work needs to be done to empirically unpack the content of those rapidly evolving representations to fully understand what they represent at different stages of processing.

**Modality-independent representations**

Our decoding approach with generalization across time and modality allowed us to investigate automatic representations of semantic categories that are independent of input modality and of low-level perceptual confounds, and to unpack their temporal evolution during production planning. We showed that modality-independent representations are automatically activated both in picture naming and overt word reading, in the absence of any category judgment task. These representations appeared to localize at occipital and temporal areas and to first evolve in parallel for both modalities but were later delayed in picture naming by 100-200ms compared to word reading. We demonstrated that those representations were not confounded by representations associated with the word forms or motor planning.

*Modality-independent representations activated at ~150ms in pictures and words*

With regards to the onsets of modality-independent representations, we found that they were activated in both picture naming and word reading as early as 100-150ms. In picture naming, modality-specific representations of semantic categories were active at 75ms. This pattern suggests that the earlier representations at 75-150ms are largely perceptually based and are followed by the later modality-independent representations starting at 150ms. This finding supports the hypothesis that initial semantic category representations with pictures are initially driven by lower-level features in the picture-stimuli while amodal representations emerge later at around 150ms. This is in line with prior evidence indicating similar onsets of modality-independent representations using stimuli of faces and places [16] and extends those findings to the context of a language production task with no explicit category judgment. Similarly, words first activated modality-specific representations around 95ms followed by amodal representations at around 150ms. The nature of the earlier, modality-specific representations remains unclear. Prior work showed that pre-lexical visual processing occurs in the first 95-180ms of the word recognition process and that graphemes and candidate lemmas are activated by 175ms [33, 43-45]. Here we showed evidence for a rapid access to semantic category as early as 95ms when reading words aloud, in line with previous work that showed ultra-rapid semantic access using spoken words [18]. Notably with words, information regarding category membership is unconfounded from then low-level visual information, therefore, one likely possibility is that the word-specific representations of semantic categories at 95-150ms represent higher-level category information, yet that they do not generalize to other modalities.

*Modality-independent representations first simultaneous for pictures and words, then delayed for pictures*

Modality-independent representations were active simultaneously in both modalities from around 100-150ms to 400ms, indicating that representations of semantic categories evolve in parallel at those times. Later on, representations at 150-400ms in words were delayed in pictures and generalized to 300-550ms. Prior work has demonstrated that similar brain activation patterns are observed early on for the perception and production of speech only to later reflect behavior-specific activity [46]. Here our results are compatible with this interpretation, reflecting the early parallel activation of shared representations that then later change in latency depending on the task. This delay in pictures could be due to the fact that pictures are generally slower to be named than words, and processes related to word retrieval and production occur faster in words compared to pictures [33]. Given the premise that word retrieval processes are related to activation of semantic information, the faster word reading process could then explain why representations that are shared in words and pictures are delayed in the picture naming pipeline compared to word reading. Finally, word-specific representations peaked at around 350ms, while modality-independent representations peaked at around 390ms in words. These times occur well after estimated times of lexical access and semantic processing [17, 19, 20], suggesting that semantic category representations are strongest once lexical selection is completed. This possibility is also further supported in our additional analyses discussed below.

All together these results indicate the modality-independent representations are activated at multiple time windows during picture naming and overt word reading, starting in parallel and then later in differ in latency in a task-specific way.

*Modality-independent vs. amodal conceptual representations: Implications for theories of concepts*

It is important to make the distinction between modality-independent representations and amodal representations. Here, it is not possible to directly assess the nature of the modality-independent representations that we have identified. On one hand, it is possible that the shared representations described here indicate amodal representations of meaning which are activated independent of the input modality. For instance, both pictures and words could activate general semantic knowledge about animals that is not necessarily grounded in perception (e.g. they are alive). This interpretation would be in line with theories of concepts which postulate the existence of an amodal conceptual hub [7]. On the other hand, an alternative could be that those shared representations encode perceptual information that gets activated similarly in words and in pictures. In other words, it is possible that both words and pictures spontaneously activate perceptual features that serve to represent semantic category information. For example, both modalities could hypothetically activate the average shape of an animal which the cross- decoding analysis would then identify as a shared representation. Although we have made efforts to keep the low-level features of the stimuli matched across s this remains a possible interpretation and would be in line with perceptual theories of concepts which argue that conceptual representations cannot be dissociated form perceptual features [1]. Nevertheless, it is very unlikely that the shared representations that we found here are caused by a reactivation of exemplar-specific perceptual features. To illustrate, visual features related to the exemplar "dog" that were activated upon encountering the picture of a dog in the experiment could be reactivated upon reading the word "dog" later in the experiment. This possibility was ruled out by our follow up analysis in which evoked responses were created by randomly sampling exemplars from each semantic category and averaging over them. By doing so we effectively averaged out exemplar-specific conceptual and visual features which could be reactivated upon encountering an exemplar a second time in a different modality. The use of a full randomization also caused the set of evoked responses in the picture modality to not match those in the word modality since both sets were created fully randomly within modalities. Thus, the only possible low-level perceptual features that could explain the cross- decoding results would be ones that survives the averaging over multiple exemplars of a semantic category and would be separable in the MEG data by the classifier. Although possible, a more likely explanation would be one where the shared representations that we have identified are indicative of amodal representations.

*Lower decoding accuracies in words vs. pictures*

Across the whole timeline and in all of our analyses, decoding accuracies were much higher for pictures compared to words with a maximum accuracy of 80% for pictures compared to 62% for words in the initial temporal decoding analysis. This can be caused by multiple factors. First, words offer a perceptually unconfounded access to semantic categories, while the low-level perceptual features in pictures are likely informing the classifier, at least in the earlier time windows. As a result, the classifier that is fit to the picture data could rely on both signal related to lower-level visual input as well as higher-order representations of categories, potentially increasing the signal-to-noise ratio and improving accuracies compared to a classifier fit to the words data. In line with that, behavioral results show that humans use early visual information to categorize pictures [47], suggesting that brain representations as early as 80ms could be informative for categorization due to biases in the visual aspect of semantic categories [9]. Second, the lower accuracy scores with the words data can partially be due to the fact that reading a word out loud does not necessarily require the activation of semantic category [22]. Participants could be relying on the mapping of letters onto phonemes and only partially activating semantic information in parallel [42]. In contrast, picture naming requires the recognition of the semantic information present in the picture, and thus necessitates the retrieval of this information to successfully recognize the target concept and to retrieve the correct lexical item to name it. As a result, the activation of category representations from words could be weaker than from pictures, in line with lower decoding scores in the word reading modality here. Despite the discrepancy in accuracy scores between picture naming and word reading, we still successfully observed reliable decoding of automatically activated category representations for both, without any explicit category judgment task.

**Visual, phonological, and lexical representations**

So far, we have shown evidence for the automatic activation of semantic category representations in picture naming and word reading. It is well established that semantic categories also interact with lexical and perceptual representations. Our additional analyses aimed at unpacking the temporal unfolding of those representations during picture naming and word reading and to contextualize them with semantic category representations.

*Semantic categories activated pre-lexically in naming, post-lexically in reading*

While on one hand, semantic category representations were shown to be activated virtually simultaneously with visual representations in picture naming, they were activated last in word reading, after visual, phonological, and lexical representations. This confirms that low-level visual representations and semantic category are highly correlated in picture naming, while in word reading semantic category is accessed after the lexical representation is reached. That is, category representations evolve from the low-level and lexical representations in words, while in pictures the visual representations in and of themselves provide information about semantic categories. This is in line with our previous findings indicating that early representations of categories in picture naming do not generalize to the word modality, and in line with previous work that indicated that category representations are automatically activated in picture naming but not necessarily in word reading [22]. This effect comes as a prediction from the Dual Route Cascaded model of reading [42] in which a word is recognized and read using both a semantic and a phonological route which operate in parallel, implying that a word can be read based on letter to sound mapping and thus does not necessarily require access to semantic and categorical representations. We thus found that with word stimuli, semantic category representations are activated only once the lexical selection process is initiated, while it occurs early with pictures and in parallel with visual representations.

*Phonological neighbors activated before lexical target in naming but simultaneously in reading*

For words, we found that visual complexity, phonological neighbors, and lexical frequency were all activated ultra-rapidly within in the first 70ms after word onset. This illustrates a quick parallel activation of different levels of representations, in line with models of reading aloud which describe the process as starting with an initial visual processing stage that is followed by lexical-semantic and phonological activation in parallel, and finally motor preparation and output [42]. In contrast, models of picture naming describe a feedforward process that starts with a visual processing stage, followed by conceptual recognition, lexical selection, phonological code retrieval, and finally motor preparation and output [33]. Our picture naming results align with those models, showing a serial unfolding of different levels of representations, with an initial activation of visual complexity, followed by phonological neighbors 25ms before lexical frequency. This demonstrates the activation of phonologically similar words prior to convergence on a target word which is in line with what is currently known about the lexical selection process [44], mainly that it operates via a spreading activation to semantically and phonologically related words, as illustrated by evidence from priming paradigms [48]. Furthermore, the difference in seriality of processing in naming compared to the parallel processing in reading illustrates that the lexical selection process which is fundamental to picture naming is absent in word reading. While picture naming required the selection of a word for production, word reading only involves a lexical recognition process, since the linguistic information is a priori provided in the input [17, 33]. Finally, we found that modality-independent lexical representations are delayed in naming compared to reading, in line with language production models which predict faster lexical access when reading words compared to naming a picture [33].

In sum, during picture naming, visual and semantic category representations are the first to be activated while lexical representations are activated last. This contrasts with the reading aloud process which starts with visual and lexical representations and ends with the activation of semantic categories. Picture naming can therefore be described as a conceptually driven process while the lexically driven word reading process seemingly goes in the opposite direction, culminating in the activation of conceptual knowledge. This suggests that conceptual knowledge can be accessed extremely early in picture naming and object recognition but that reading requires the target word to be recognized before semantic knowledge is accessed. The investigation of concepts using conceptual statistics and experiential features has brought an important contribution to the field [3, 11], allowing to go beyond the investigation of concepts at the semantic category level, and opening the possibility to assess the neural basis of concepts at a finer granularity. Future work would have to go beyond the superordinate category level and assess our findings generalize to different hierarchies of conceptual knowledge (e.g., basic level, subordinate level). By doing so, we would hope to get closer to unpacking the mechanism by which words map on to concepts which is at the core of the field.

**CONCLUSION**

We found evidence for the automatic activation of modality-independent representations of semantic categories from around 150ms to 500ms for both picture naming and word reading, and in the absence of an explicit categorization task. Those representation were shown to first evolve in parallel for both tasks, and then be delayed in naming. The activation of modality-specific representations early followed by later amodal representations suggests a specific-to-generic activation of semantic representations. Finally, semantic category was shown to be activated before lexical access is initiated for pictures but after lexical access for words, reflecting a conceptually driven process in picture naming which contrasts with the lexically driven process in word reading. Future work should investigate modality-independent representations at finer granularities that go beyond the superordinate category level.

Acknowledgements

This work was supported by the NYUAD Research Institute under Grant G1001.